\documentclass[11pt]{amsart}
\usepackage{amsmath,amssymb,amsfonts}
\usepackage{graphicx}
\usepackage{hyperref}
\hypersetup{hidelinks=true}

\usepackage{subcaption}    
\usepackage{caption}       
\usepackage{tabularx}      
\usepackage{multirow}      
\usepackage{booktabs}      
\usepackage{array}         
\usepackage{float}         
\usepackage{placeins}      
\usepackage{xcolor}        
\usepackage{makecell}      

\usepackage{algorithmic}   
\usepackage{algorithm}     
\usepackage{textcomp}
\usepackage{stfloats}
\usepackage{url}
\usepackage{verbatim}
\usepackage{cite}

\DeclareMathOperator{\Birth}{Birth}
\DeclareMathOperator{\Death}{Death}
\DeclareMathOperator{\Lifetime}{Lifetime}

\title[Integrating TDA and DNN for GPR Data] {Shape-Aware Topological Representation for Pipeline Hyperbola Detection in GPR Data}

\author[M. Kang]{Meiyan Kang}
\address{Department of mathematics, Ajou University, 206, World cup-ro, Yeongtong-gu, Suwon 16499,  Republic of Korea}
\email{miyeon@ajou.ac.kr}

\author[S. Kaji]{Shizuo Kaji}
\address{Institute of Mathematics for Industry, Kyushu University, Fukuoka, 819-0395, Japan; Graduate School of Science, Kyoto University, Kyoto, 606-8502, Japan}
\email{skaji@imi.kyushu-u.ac.jp}

\author[S.-Y. Lee]{Sang-Yun Lee}
\address{Construction Environment System Research Institute, Inha University, Incheon, 22212, Republic of Korea}
\email{sangyunlee@kict.re.kr}

\author[T. Kim]{Taegon Kim}
\address{Department of mathematics, Ajou University, 206, World cup-ro, Yeongtong-gu, Suwon 16499,  Republic of Korea}
\email{nayo6180@gmail.com}

\author[H.-H. Ryu]{Hee-Hwan Ryu}
\address{Transmission \& Substation Laboratory, Korea Electric Power Corporation (KEPCO) Research Institute, Daejeon, 34056, Republic of Korea}
\email{hhryu82@kepco.co.kr}

\author[S. Choi]{Suyoung Choi}
\address{Department of mathematics, Ajou University, 206, World cup-ro, Yeongtong-gu, Suwon 16499,  Republic of Korea}
\email{schoi@ajou.ac.kr}

\date{\today}


\thanks{This research was supported by Basic Science Research Program through the National Research Foundation of Korea(NRF) (RS-2021-NR060141) and Korea Electric Power Corporation (R21SA02). The second author is partially supported by JST Moonshot R\&D Grant Number JPMJMS2021 and the commissioned research (No.22301) by NICT, Japan.}

\begin{document}

\begin{abstract}
Ground Penetrating Radar (GPR) is a widely used non-destructive testing (NDT) technique for subsurface exploration, particularly in infrastructure inspection. 
However, traditional interpretation methods often struggle with noise sensitivity and limited structural awareness.
We propose a novel framework that integrates shape-aware topological features, derived from B-scan GPR images using Topological Data Analysis (TDA), with the object detection capabilities of a YOLOv5-based deep neural network (DNN). 
This topological representation improves geometric salience, enhancing detection and localization of underground utilities, especially pipelines.
To mitigate the scarcity of annotated real-world data, a Sim2Real strategy is employed.
Synthetic datasets are generated to capture both diverse subsurface conditions and the essential hyperbolic reflection patterns of pipelines, enabling more effective knowledge transfer to real-world scenarios.
Experimental results show consistent improvements in mean Average Precision (mAP), highlighting the robustness and effectiveness of the proposed method. This work demonstrates the potential of TDA-enhanced deep learning for reliable subsurface object detection with broad implications in urban planning, safety inspection, and infrastructure management.
\end{abstract}

\maketitle

\section{Introduction}

Inspection of subsurface infrastructure--particularly pipelines--is critical for maintaining urban environments and preventing failures~\cite{yang2023adaptive}. This task necessitates high-accuracy non-destructive testing (NDT) methods~\cite{neal2004ground}. 
Among these, \emph{Ground Penetrating Radar (GPR)} has emerged as a widely adopted geophysical technique, facilitating the detection and imaging of subsurface structures across diverse domains such as archaeology, civil engineering, and environmental studies~\cite{daniels2004ground, jol2008ground}. 
GPR operates by transmitting electromagnetic waves into the ground and interpreting the resulting reflections to infer subsurface compositions.
Traditionally, the interpretation of GPR signals has relied heavily on manual analysis and physics-based models. 
These conventional approaches are not only time-consuming and labor-intensive, but also highly susceptible to noise and environmental variability. 
Such limitations have motivated the exploration of automated solutions.

Recent advances in deep learning have driven a paradigm shift in subsurface sensing tasks~\cite{hong2023intelligent, su2022prediction, lee2024void, liu2021efficient, ryu2025machine}. 
Similarly, in Ground Penetrating Radar (GPR), deep learning techniques have been applied to enable significant improvements in automation, detection accuracy, and computational efficiency.
Notably, object detection models--including YOLO (You Only Look Once)~\cite{redmon2016}, Faster R-CNN~\cite{ren2016faster}, and U-Net~\cite{ronneberger2015}--have demonstrated considerable promise. 
YOLO excels in rapid classification of subsurface anomalies (e.g., pipelines, voids), requiring minimal post-processing. 
In contrast, Faster R-CNN often achieves higher accuracy, especially for small or low-contrast objects, but at the cost of increased computational demands, though it typically incurs greater computational cost~\cite{Fang_2021GPR}. 
U-Net is particularly effective for pixel-level segmentation, aiding in the delineation of complex subsurface structures, though it is prone to overfitting when annotated data is scarce.

Beyond individual model architectures, recent research has explored integrated frameworks for GPR interpretation. For example, Lei et al.~\cite{lei2024gpr} combined deep learning with reverse time migration to improve localization accuracy. 
Similarly, Su et al.~\cite{su2023} introduced an end-to-end deep learning model for underground utility detection, while Xiong et al.~\cite{xiong2023gprgan} proposed GPR-GAN to augment training via synthetic data generation.
In addition, Jafuno et al.~\cite{jafuno2024secondorder} applied second-order deep learning models to enhance buried object classification, and Tag et al.~\cite{tag2024yolov8gpr} demonstrated effective groundwater detection using YOLOv8 on GPR datasets.

Despite these advances, a key limitation persists: the performance of deep learning models is heavily reliant on the availability and diversity of high-quality annotated data.
Collecting labeled GPR datasets is expensive and logistically challenging, as field measurements are often limited by access, cost, and environmental constraints.
Recent studies have explicitly highlighted the scarcity of publicly available GPR datasets for underground utility detection, emphasizing that the lack of large-scale, standardized field data remains a critical bottleneck for deep-learning-based interpretation~\cite{Mojahid2025Intelligent,Paul2025Review}.
This motivates the increased adoption of simulation-based training and domain adaptation strategies.

\emph{Sim-to-Real (Sim2Real)} transfer has emerged as a viable solution by enabling models trained on simulated GPR data to generalize to real-world environments~\cite{JUNKAI2024105333}.
Simulated environments allow for the generation of diverse, labeled, and noise-controlled data representing various subsurface conditions.
These approaches help bridge the domain gap between synthetic and field data, improving model generalization and robustness.

Another promising direction lies in improving feature representation.
\emph{Topological Data Analysis (TDA)} provides a mathematical framework that captures intrinsic structural patterns often overlooked by conventional methods ~\cite{edelsbrunner2010computational, ghrist2008barcodes}.
TDA is particularly appealing in GPR analysis due to its invariance to coordinate transformations and robustness to noise~\cite{bubenik2015, wasserman2018topological}.
Its core technique, persistent homology, encodes the birth and death of topological features across multiple scales, producing a compact yet expressive representation of data geometry~\cite{chazal2021}. 

While previous studies have primarily utilized vectorized persistent homology such as barcodes or persistence diagrams as features for machine learning, this work takes a novel approach: we directly integrate topological features derived from persistent homology with raw B-scan GPR data. This fusion enables the model to learn both spatial textures and underlying shape structures, resulting in richer representations and improved detection performance.

Unlike conventional TDA–CNN approaches that employ statistical embeddings such as persistence images (PI)~\cite{adams2017persistence}, persistence entropy (PE)~\cite{chintakunta2015entropy}, or persistence landscapes (PL)~\cite{bubenik2015landscapes},
which summarize persistence diagrams into abstract feature vectors, our method preserves spatial coherence through a \textit{shape-aware topological representation}. 
This representation directly encodes the geometric continuity and connectivity of reflection hyperbolas within the image domain, 
allowing topological information to be physically aligned with pixel-level GPR reflections. 
Consequently, it bridges the gap between topological abstraction and geometric interpretability, providing a more intuitive and robust integration of TDA into deep-learning-based GPR analysis.

In this paper, we propose a novel framework, \emph{TE-S2R} (TDA-Enhanced Sim2Real), for robust subsurface object detection. The core idea is to augment conventional deep learning pipelines with shape-aware topological features extracted via TDA, which effectively capture structural information while remaining resilient to noise. Unlike prior work, our approach directly incorporates topological features as enhanced image representations, enabling the model to learn both geometric and spatial characteristics more effectively.

A key challenge in Sim2Real approaches lies not in reducing data collection costs, but in bridging the distributional gap between synthetic and real-world data, particularly with respect to structural inconsistencies and noise. TE-S2R addresses this issue by integrating TDA-based features that emphasize shape and connectivity over raw pixel values. This reduces sensitivity to domain-specific noise and enhances topological alignment between synthetic and real GPR images. As a result, our method facilitates more stable and reliable knowledge transfer across domains. The synergy between TDA and Sim2Real yields a significant performance boost under real-world conditions, where noise and variability are unavoidable.

The resulting TE-S2R framework demonstrates consistent improvements in mean Average Precision (mAP), offering a scalable and noise-resistant solution that is effective not only for GPR data, but also extensible to other types of grid-based imaging modalities.

\section{Methodology} \label{sec:methodology}

The aim of this study is to accurately detect underground pipelines using GPR. In particular, we focus on analyzing B-scan images, which are the primary output of GPR measurements. 
A B-scan represents a two-dimensional cross-sectional image generated by sweeping the radar antenna along a linear path.
While B-scan images provide valuable information about subsurface structures, they are often contaminated with noise and clutter, making automated interpretation a challenging task. 
To address this issue, we propose utilizing TDA to enhance detection performance by extracting robust, noise-resistant features that capture the inherent shape of the data.
Specifically, we capture topological structures embedded in B-scan images, such as loops formed by hyperbolic reflections from buried cylindrical objects. These shape-aware topological features are then fused with the original image data, resulting in enriched structural representations that enhance the model's sensitivity to object geometry.

\subsection{Step 1: Shape-Aware Topological Representation} \label{sec:method_step1_tda}

\emph{Persistent homology} is a computational technique within TDA that extracts multi-scale structural features from 2D digital images by analyzing how topological properties evolve across different threshold levels~\cite{carlsson2009topology}.

In the context of 2D image analysis, persistent homology operates on \emph{cubical complexes} constructed directly from pixel grids. A cubical complex for a 2D image is built by treating each pixel as a 2-dimensional cube (square), with edges and vertices forming the 1-dimensional and 0-dimensional cubes respectively~\cite{Kaczynski-Mischaikow-Mrozek2004book}. This representation is particularly natural for digital images since it preserves the inherent grid structure without requiring triangulation or point cloud conversion.

\begin{figure*}[thbp]
    \centering
    \subfloat[Grayscale image]{
        \includegraphics[width=0.35\textwidth]{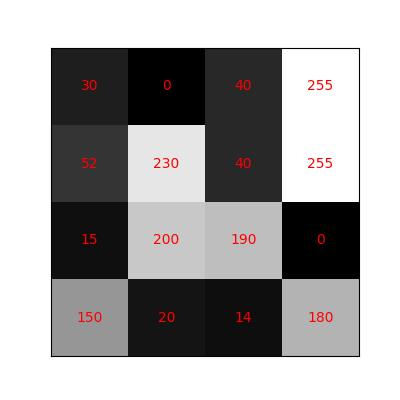}
    }
    \subfloat[Filtration.]{
        \includegraphics[width=0.60\textwidth]{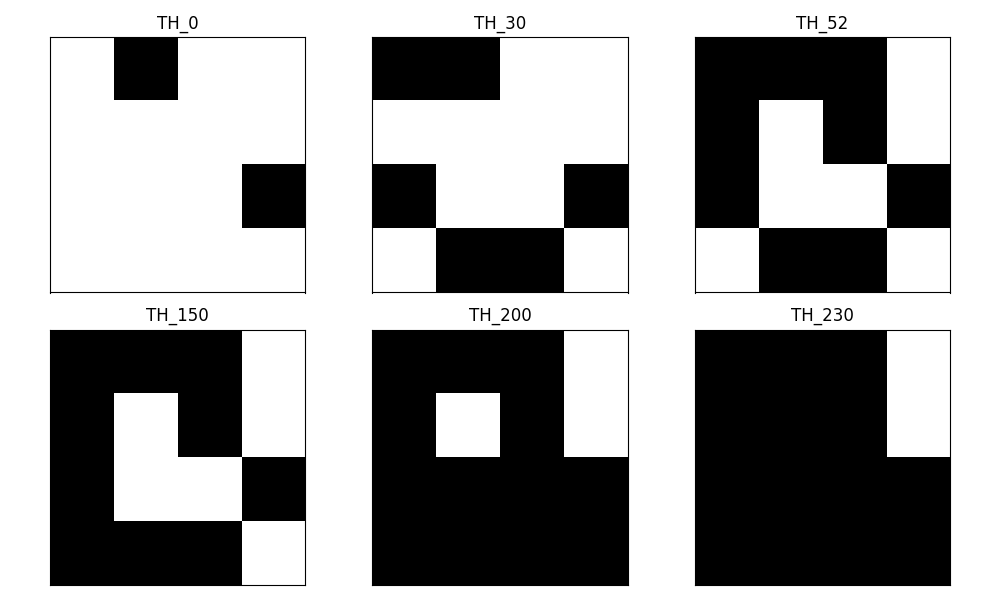}
    }   
    \caption{
        Persistent homology applied to grayscale imagery. (a) A typical grayscale image used for topological analysis. (b) Filtration process for constructing a filtered cubical complex through intensity-based thresholding. As the threshold increases from 0 to 255, pixels are progressively included in the complex.
    }\label{fig:filtration_example}
\end{figure*}

The persistent homology computation on 2D images typically proceeds through a \textit{filtration} process based on pixel intensity values. Starting from an empty complex, cubes (pixels) are incrementally added in order of increasing (or decreasing) intensity values, creating a nested sequence of cubical complexes. As this filtration progresses, topological features emerge and vanish: connected components appear when pixels are first added (\(\beta_0\) features), while loops or holes form when pixel arrangements create enclosed regions (\(\beta_1\) features).
See Figure~\ref{fig:filtration_example} for illustration.

For each topological feature \(\sigma\) detected during this process, persistent homology records its \textit{birth} time (the intensity value when the feature first appears) and \textit{death} time (when it disappears due to subsequent pixel additions). The \emph{lifetime} of the feature is computed as:
\begin{equation}
    \Lifetime(\sigma) = \Death(\sigma) - \Birth(\sigma)
    \label{eq:lifetime}
\end{equation}
In 2D image analysis applications, \(\beta_0\) features (connected components) capture blob-like structures and their hierarchical merging behavior across intensity scales, while \(\beta_1\) features (loops) identify ring-like or hole structures within the image. Features with long lifetimes typically correspond to significant image structures, whereas short-lived features often represent noise or minor intensity fluctuations. This multi-scale analysis enables robust feature extraction that is less sensitive to noise compared to traditional computer vision approaches, making persistent homology particularly valuable for medical imaging, texture analysis, and shape recognition tasks where topological structure carries important semantic information~\cite{brito2025persistent}.

The core contribution of this study lies in bridging the gap between topological data analysis and convolutional neural networks (CNNs) for enhanced shape-aware object detection. 
While CNNs excel at capturing local textural patterns, they are inherently biased toward low-level features and struggle to capture global shape information. 
This limitation is particularly problematic for detecting buried pipelines in GPR imagery, where target objects manifest as distinctive curved, hyperbolic patterns that require global shape understanding rather than local texture analysis.

To address this fundamental limitation, we develop a novel method that enables CNNs to effectively digest topological and global shape information by encoding persistent homology-derived features directly into image representations. 
Our approach transforms abstract topological annotations into a visual format that CNNs can naturally process, thereby augmenting their shape perception capabilities without requiring architectural modifications.

We treat each B-scan image as a grayscale 2D image and apply the persistent homology framework.
We focus exclusively on $\beta_1$ topological features (loops), as our primary interest lies in the crescent-shaped closed patterns representing subsurface pipelines. 

The key idea of our approach lies in transforming abstract topological information into a format that CNNs can effectively utilize. 
Rather than relying on conventional persistence diagrams or barcodes, which provide compact summaries but cannot be directly processed by CNNs, we construct a spatial representation that preserves both topological significance and spatial locality.
We create a new image with identical dimensions to the original input, visualizing the shapes of degree-one homology generators while modulating their intensity according to their lifetime. 
Generators with longer lifetimes are rendered with higher intensity, while shorter-lived features appear with lighter shading. 


The detailed procedure for constructing the lifetime-weighted topological feature map 
is summarized in Algorithm~\ref{alg:lifetime_map}. 
Each GPR B-scan is first converted into a grayscale matrix and processed through a cubical filtration 
to compute persistent homology using the \texttt{Ripserer.jl} package implemented in the \texttt{Julia} environment. 
The resulting $H_1$ generators representing loop-like topological structures are extracted along with their birth and death times, 
and their normalized lifetimes $(d_i - b_i)$ are used as transparency weights~$\alpha_i$ to indicate topological significance. 
These generators are then visualized on an image grid and their weighted intensities are accumulated onto the original B-scan domain, 
producing a spatially aligned and topologically enhanced representation~$T$ that preserves both amplitude and geometric stability. 
This fused map serves as the final input to the CNN backbone without requiring any architectural modification. 

\begin{algorithm}[tpbp]
\caption{Generation of Lifetime-Weighted Topological Feature Map}
\label{alg:lifetime_map}
\begin{algorithmic}[1]   
\STATE \textbf{Input:} B-scan image $I$
\STATE \textbf{Output:} Lifetime-weighted feature map $T$

\STATE Convert $I$ to grayscale matrix $G \in [0,1]$
\STATE Reverse $G$ vertically to align coordinate orientation
\STATE Compute persistent homology using cubical filtration:
       $PH \leftarrow \texttt{ripserer(Cubical($G$); cutoff=0.2)}$
\STATE Extract $H_1$ generators with representative cycles from $PH$
\IF{$H_1 \neq \emptyset$}
    \FOR{each generator $g_i \in H_1$}
        \STATE Compute lifetime $l_i = d_i - b_i$
    \ENDFOR
    \STATE Normalize $l_i$ to $[0.3, 1.0]$ as transparency weights $\alpha_i$
    \STATE Initialize $T \leftarrow \mathbf{0}$ with the same size as $G$
    \FOR{each generator $g_i \in H_1$}
        \FOR{each edge $(v_1, v_2)$ in representative$(g_i)$}
            \STATE Draw a line between $(v_1, v_2)$ on $H$
                   with color = green, opacity = $\alpha_i$
            \STATE Accumulate $\alpha_i$ into $T$ at corresponding coordinates
        \ENDFOR
    \ENDFOR
\ELSE
    \STATE Skip visualization (no $H_1$ features detected)
\ENDIF
\STATE Save the overlay visualization $H$ and feature map $T$
\end{algorithmic}
\end{algorithm}

This spatial encoding enables CNNs to simultaneously access local pixel information and global topological structure within their standard convolution operations.
To construct the final input for the object detection model, we stack the grayscale GPR B-scan image and the lifetime-weighted topological feature map along the channel dimension, forming a multi-channel representation. 
This approach allows the CNN-based detector (YOLOv5) to learn from both raw spatial information and topological cues, effectively augmenting its shape perception capabilities.
The CNN can thus leverage its existing architectural strengths for local feature detection while gaining access to global topological information that would otherwise be inaccessible through standard convolution operations.

\begin{figure*}[thbp]
    \centering
    \subfloat[Original B-scan GPR image]{
        \includegraphics[width=0.31\linewidth]{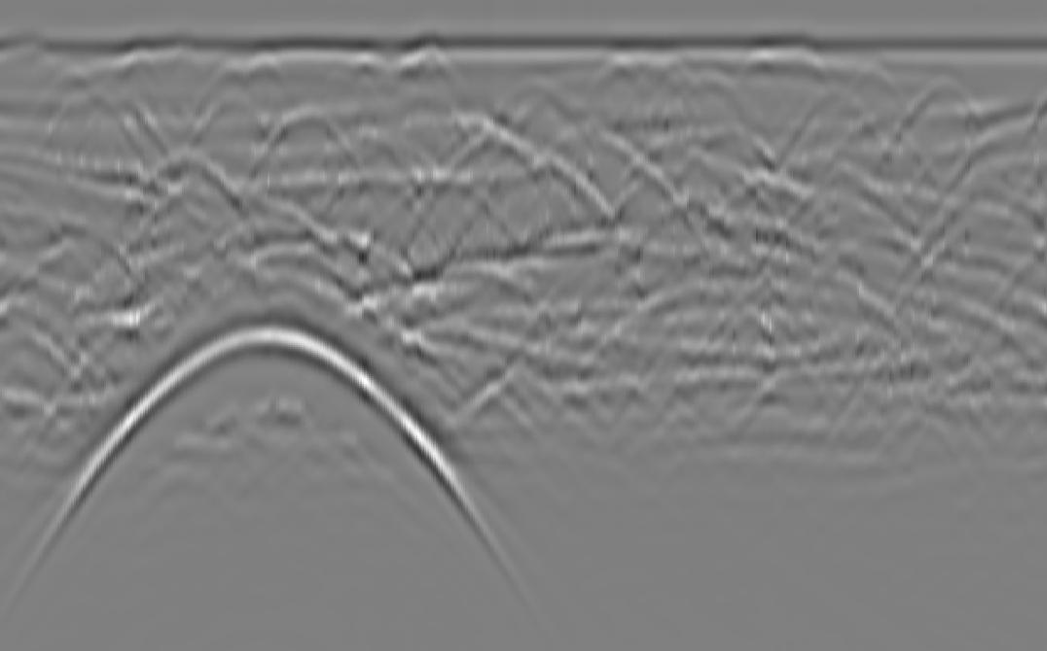}
    }
        \hfill
    \subfloat[Persistent homological features]{
        \includegraphics[width=0.31\linewidth]{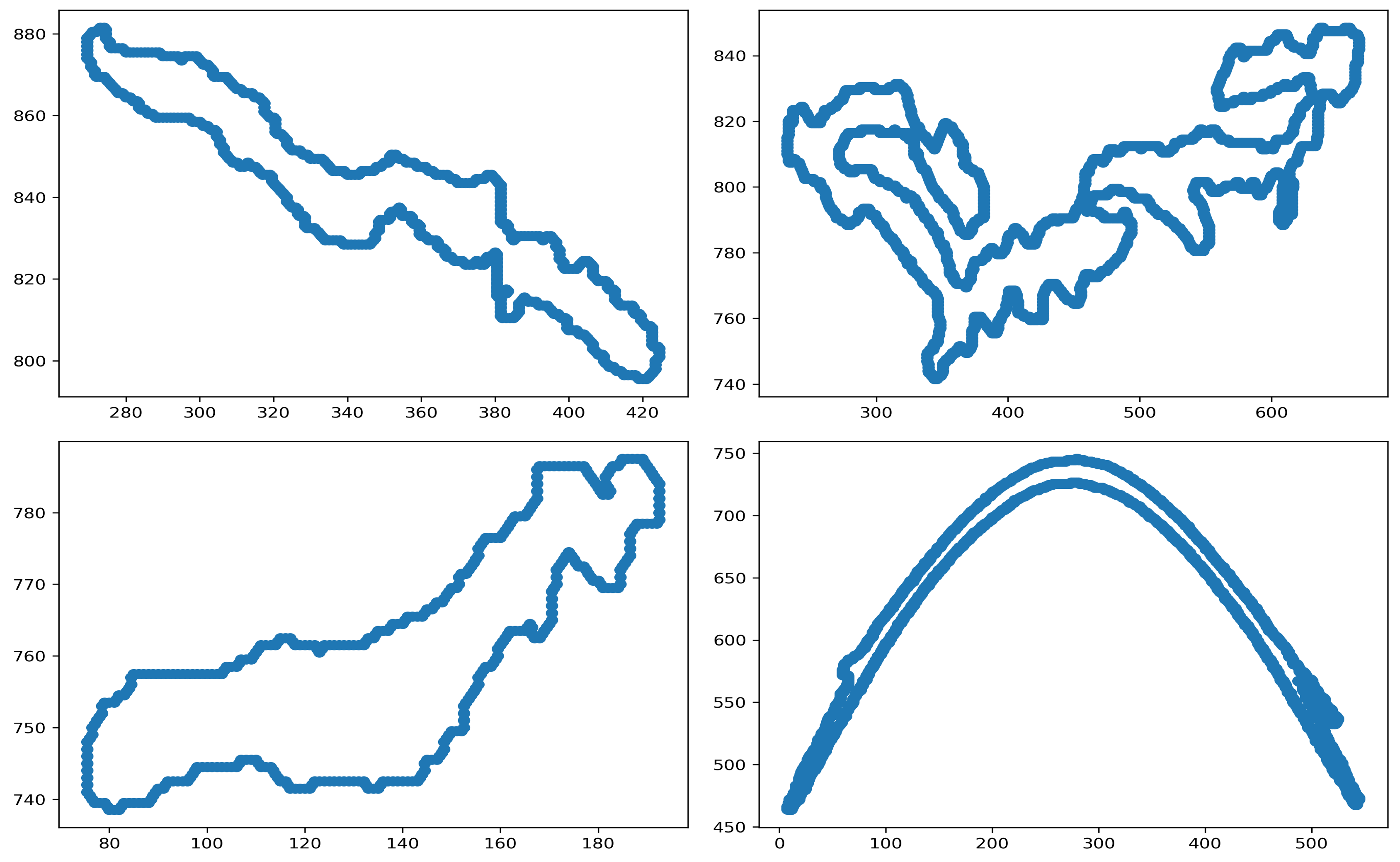}
    } 
        \hfill
    \subfloat[Shape-aware representation]{
        \includegraphics[width=0.31\linewidth]{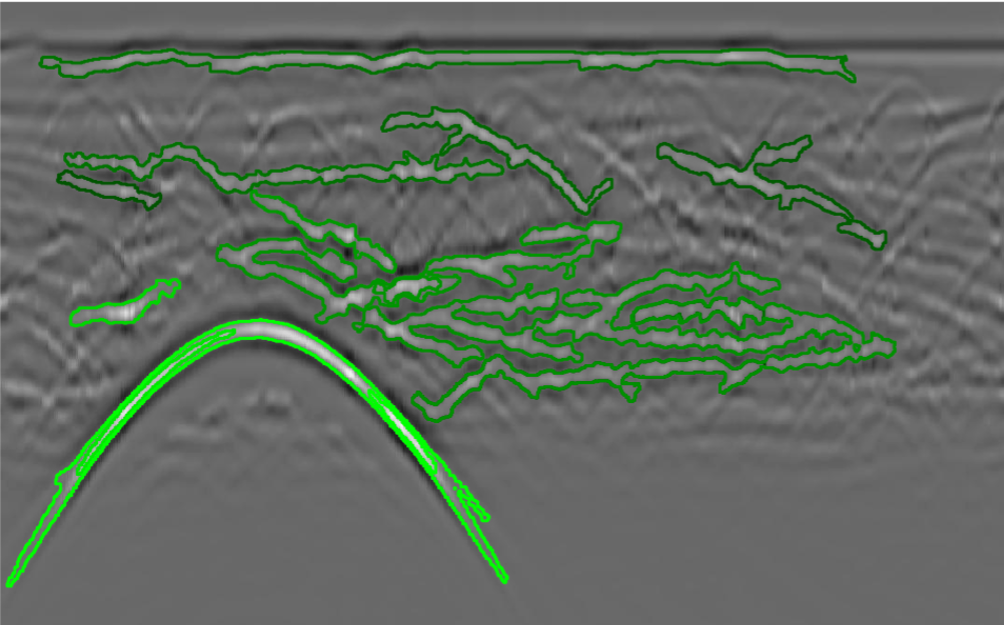}
    }    
    \caption{
    Topological feature construction from a GPR image. (a) Original GPR B-scan. (b) Persistent homological features extracted via $H_1$ generators. (c) Shape-aware representation formed by stacking the original image and persistence-weighted features along the channel dimension.
    }
    \label{fig:tda_example}
\end{figure*}

This represents a novel contribution that enables CNNs to overcome their inherent shape-blindness by providing explicit topological guidance encoded in a spatially coherent manner. 
To the best of our knowledge, this approach of encoding topological lifetime information as spatial intensity maps for CNN consumption has not been addressed in prior research.
An example of such a visualization, constructed from real GPR B-scan data using the proposed method, is shown in Fig.~\ref{fig:tda_example}.

\begin{figure*}[thbp]
    \centering
    \includegraphics[width=\linewidth]{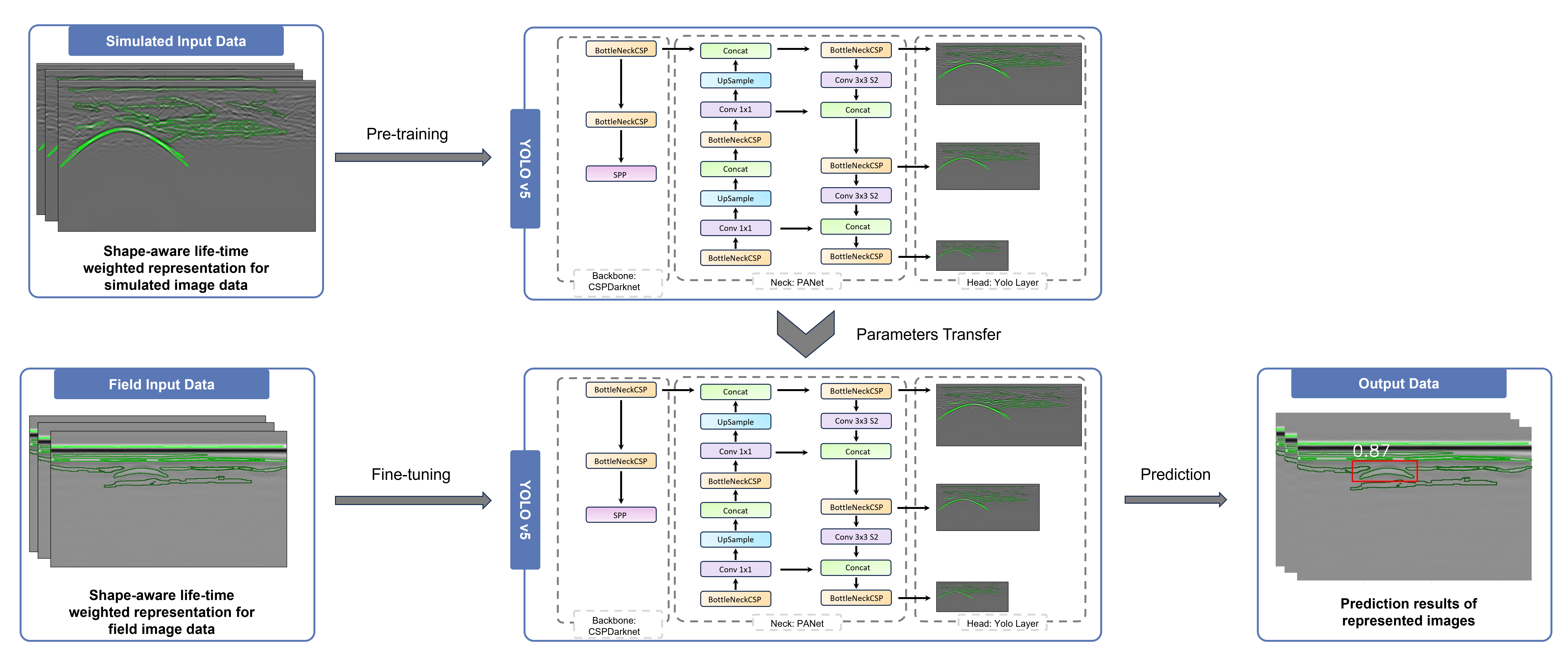}
    \caption{
    Sim2Real training process for GPR detection. The model is first pre-trained on simulated data and then fine-tuned on field data. In both stages, inputs are transformed into shape-aware representations by combining the original GPR images with persistent homology-derived topological features.
    }
    \label{fig:sim2real_process}
\end{figure*}

\subsection{Step 2: Sim2Real Training}

Collecting and annotating real-world GPR data presents significant challenges, as subsurface environments are inherently complex and buried utilities are difficult to access for accurate ground-truth labeling. 
This labour-intensive annotation process makes training YOLOv5 directly on large-scale field datasets impractical. 

To overcome this data scarcity challenge, we adopt a Sim2Real training philosophy that leverages synthetic data generation followed by real-world fine-tuning. 
Our approach begins with pre-training the model on a comprehensive dataset of synthetic GPR B-scan images generated using \emph
{gprMax}, with the data generation procedure detailed in Section IV.~\ref{sec:numerical_data_generation}. 
These simulated images undergo processing through our TDA pipeline (Section~\ref{sec:method_step1_tda}), where $\beta_1$ topological features corresponding to hyperbolic reflections are extracted and encoded into lifetime-weighted shape-aware representations.

This synthetic pre-training phase enables the model to acquire fundamental knowledge of subsurface reflection patterns and their topological signatures without requiring extensive real-world annotations. 
The enriched synthetic data, augmented with persistent homology features, provides a robust foundation for learning the characteristic shapes and structures associated with buried pipelines.

Following the pre-training phase, the model is fine-tuned using a curated set of real-world GPR scans obtained from a controlled testbed environment.
These field data exhibit complex variations due to factors such as soil heterogeneity, depth-dependent signal attenuation, and sensor-induced noise. 
The field data acquisition and experimental setup are detailed in Section~\ref{sec:field_data_collection}.

Fine-tuning on field data is essential for bridging the domain gap between the idealized conditions of synthetic simulations and the stochastic, site-specific nature of real-world measurements. 
While pre-training enables the model to learn generalized geometric and topological priors, fine-tuning refines the learned parameters to adapt to real-world complexity.
This process enhances the model’s robustness and adaptability, allowing it to detect nuanced subsurface features more effectively. 
By incorporating a diverse range of field samples, the fine-tuning phase significantly improves real-world performance, enabling reliable detection of buried utilities under challenging and variable environmental conditions.

This two-stage approach maximizes the utility of limited real-world data while leveraging the scalability of synthetic data generation, ultimately achieving robust performance in practical deployment scenarios.
The complete training schema is illustrated in Fig. ~\ref{fig:sim2real_process}.

\section{Experiments and Results}\label{sec:experiments}

\subsection{Numerical Data Generation} \label{sec:numerical_data_generation}

We employed the Finite-Difference Time-Domain (FDTD) method, which discretizes time and space to numerically solve Maxwell's equations, enabling accurate simulation of electromagnetic wave propagation.
For this purpose, we used the open-source software gprMax, which allows efficient modeling of complex subsurface environments~\cite{warren2016gprmax}.

The simulation setup included essential components such as pipe placement, soil layering, and grid resolution to reflect realistic underground environments. 
A Ricker wavelet--a zero-phase wavelet commonly used in GPR applications due to its compact time-domain representation--with a 350MHz center frequency was employed as the excitation signal. 
The radar system was modelled with a co-located transmitter–receiver pair that moved along the surface to acquire successive A-scans forming each B-scan cross-sectional image.

To specify ground material properties accurately, we adopted the Peplinski model~\cite{peplinski1995dielectric}, which provides empirical relationships for calculating the complex permittivity of soils based on physical parameters such as soil moisture content, bulk density, and particle composition. For enhanced realism, heterogeneous soil conditions were incorporated using a fractal-based stochastic generation algorithm~\cite{molz2004stochastic}, which spatially distributes material properties in a self-similar pattern that mimics natural soil variability.
In contrast, homogeneous scenarios assumed constant permittivity values based on averaged water content across the simulation domain.

The specific input parameters used for the Peplinski and fractal models are summarized in Table~\ref{tab:peplinski}.
These include the sand and clay content, bulk and particle density, and the range of volumetric water content, as well as the fractal dimension ($D_\text{frac}$) and directional weights ($W_\text{frac}$). 
These parameters were selected within realistic ranges to generate diverse dielectric and conductive profiles for pretraining the Sim2Real transfer-learning model, rather than to reproduce site-specific soil conditions.
This approach follows the common practice in GPR simulation studies, where synthetic data are primarily used to initialize deep-learning models before real-data adaptation~\cite{Raha2024}.

\begin{figure*}[thbp]
    \centering
    \subfloat[Schematic diagram of the simulation model]{
        \includegraphics[width=0.45\textwidth]{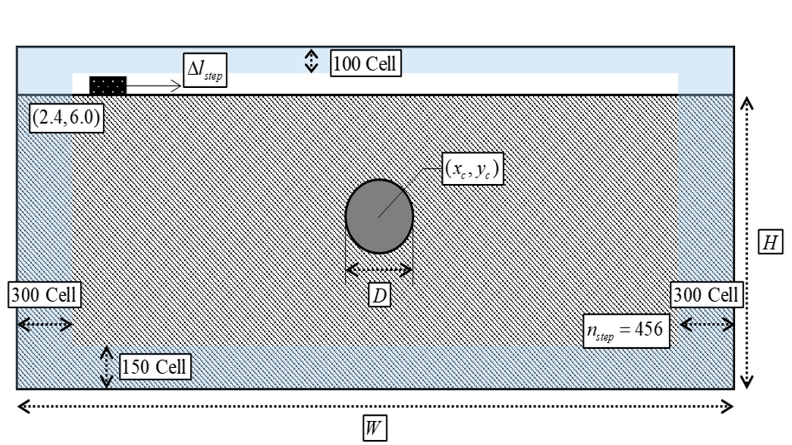}
    }
    \subfloat[On-site data collection in the field experiment]{
        \includegraphics[width=0.40\textwidth]{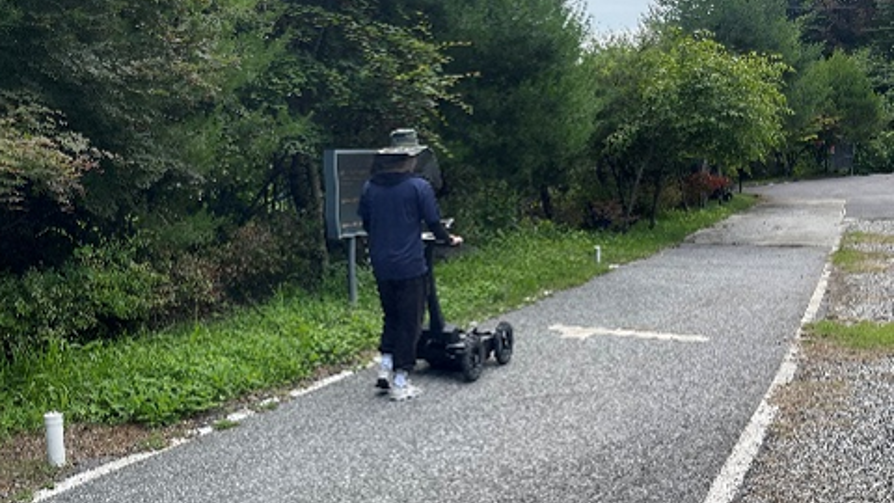}
    }
    
    \subfloat[Simulated B-scan image]{
        \includegraphics[width=0.40\textwidth]{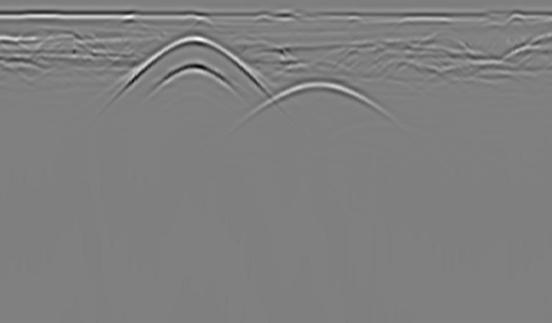}
    }\hspace{8mm}
    \subfloat[Real-world field B-scan image]{
        \includegraphics[width=0.40\textwidth]{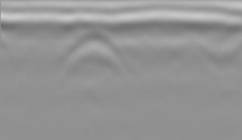}
    }

    \caption{
     Overview of the experimental setup and representative data samples.  
    (a) Schematic diagram of the simulation model showing the pipe geometry, grid configuration, and boundary conditions, with relevant dimensions labeled (pipe diameter = 0.5~m, burial depths = 0.3--0.4~m, grid resolution = 6~mm).  
    (b) On-site photograph of the field data collection setup showing the GPR scanning apparatus and buried pipe layout.  
    (c) Example of a simulated B-scan image generated from the synthetic model corresponding to the setup in (a).  
    (d) Example of a real B-scan image collected from Site~A, corresponding to a 2~m survey line and a buried pipe at a depth of approximately 0.5~m.  
    }
    \label{fig:bscan_results}
\end{figure*}

The 2D simulation setup is illustrated in Fig. ~\ref{fig:bscan_results}(a).
The transmitter and receiver antennas were positioned 12~cm apart and moved together along the survey direction in 0.024~m increments per A-scan, resulting in 456 traces per B-scan.
Both antennas were placed directly on the ground surface without a separation gap, while the air layer above the surface was modeled with standard air properties.

\begin{table}[thbp]
\caption{Peplinski and Fractal Input Parameters Used in the FDTD Simulation}
\label{tab:peplinski}
\centering
\begin{tabular}{lccccc}
\toprule
\textbf{Model} & $S$ (\%) & $C$ (\%) & $\rho_b$ (g/cm$^3$) & $\rho_s$ (g/cm$^3$) & $w_{\text{range}}$ \\
\midrule
Peplinski & 80 & 20 & 2.0 & 2.66 & [0.01–0.10] \\
\midrule
\textbf{Fractal} & $N_{\text{mat}}$ & $D_{\text{frac}}$ & \multicolumn{3}{c}{$W_{\text{frac}}$ [x, y, z]} \\
\midrule
& 10 & 1.0 & \multicolumn{3}{c}{[2.0, 0.1, 0.1]} \\
\bottomrule
\end{tabular}
\end{table}

Table~\ref{tab:model_params} summarizes the key parameters used in the simulation.
Pipe diameters ranged from $0.3\mathrm{m}$ to $1.0\mathrm{m}$, with horizontal positions between $5.0\mathrm{m}$ and $11.0\mathrm{m}$, and burial depths ranging between $3.5\mathrm{m}$ and $5.3\mathrm{m}$.
Absorbing boundary conditions were applied on all sides with cell padding of $[300, 300, 100, 150]$, corresponding to $1.8,\mathrm{m}$ (left and right), $0.6,\mathrm{m}$ (top), and $0.9,\mathrm{m}$ (bottom), based on the grid resolution of $6,\mathrm{mm}$.

B-scan images were generated through 2D electromagnetic simulations under both homogeneous and heterogeneous conditions, with scenarios with single and multiple pipes.To improve the quality and interpretability of the synthetic B-scans, 
standard GPR preprocessing operations (background removal, band-pass filtering, and automatic gain control) 
were applied, consistent with the procedures used for real data (see Section~\ref{sec:field_data_collection}).

For each B-scan image, five AGC variants were generated to increase the model’s robustness against intensity variations. In total, 300 simulated B-scan images were produced. An example output is shown in Fig. ~\ref{fig:bscan_results}(c).

It is important to note that the simulated data were not generated to replicate any specific field site. 
Instead, the simulation stage was designed to facilitate the learning of fundamental GPR reflection patterns (e.g., hyperbolic responses) under controlled conditions. 
The real-world datasets (Sites~A, B, and~C) were independently collected from distinct geographical areas with different soil types and noise characteristics. 
Hence, the proposed Sim2Real framework focuses on transferring generalizable feature representations from synthetic to heterogeneous real-world environments, 
rather than modeling a single site-specific scenario.

\begin{table}[htbp]
    \centering
    \caption{Modeling input properties for gprMax simulations}
    \label{tab:model_params}
    \begin{tabularx}{\columnwidth}{l m{3.5cm} >{\raggedright\arraybackslash}X}
        \toprule
        \textbf{Category} & \textbf{Parameter} & \textbf{Value} \\
        \midrule
        \multirow{4}{*}{Transmission} 
            & Wave type & Ricker \\
            \cmidrule(lr){2-3}
            & Center frequency ($f$) & 350~MHz \\
            \cmidrule(lr){2-3}
            & A-Scan steps ($n_{\text{step}}$) & 456 \\
            \cmidrule(lr){2-3}
            & Step-wise movement ($\Delta l_{\text{step}}$) & 0.024~m/step \\
        \midrule
        \multirow{3}{*}{Model Geometry} 
            & Model size ($W \times H$) & 16.0m $\times$ 6.0~m \\
            \cmidrule(lr){2-3}
            & Pipe diameter ($D$) & [0.3, 0.5, 1.0]~m \\
            \cmidrule(lr){2-3}
            & Pipe center ($x_c, y_c$) & ([5.0–11.0]~m, [3.5–5.3]~m) \\
        \midrule
        Grid & Cell size ($\Delta l$) & 6mm \\
        \midrule
        Boundary & Absorbing BC ($BC$) & [1.8, 1.8, 0.6, 0.9]~m \\
        \bottomrule
    \end{tabularx}
\end{table}

\subsection{Field Data Collection}
\label{sec:field_data_collection}

To evaluate the proposed method under realistic subsurface conditions, GPR B-scan data were collected from three field sites in South Korea:

\begin{itemize}
    \item \textbf{Site A}: 451 Daragjae-ro , Seorak-myeon, Gapyeong-gun, Gyeonggi-do, Republic of Korea
    \item \textbf{Site B}: 2, Busandaehak-ro 63beon-gil, Geumjeong-gu, Busan, Republic of Korea
    \item \textbf{Site C}: 341, Baekbeom-ro, Yongsan-gu, Seoul, Republic of Korea
\end{itemize}

Each site contains pipes installed at known depths and orientations under heterogeneous soil conditions. 
Scans were acquired along predefined linear paths using a commercial GPR system developed by Systems, Inc. (USA). 
The system comprises a main control unit (\emph{SIR 4000}), a $350$~MHz antenna (\emph{Model 350HS}), and a survey cart with an encoder wheel (\emph{Model 653}) for accurate distance tracking. 
This configuration offers a practical trade-off between resolution and penetration depth, and enables consistent data acquisition across sites.

According to the manufacturer’s documentation~\cite{GSSI2017manual,GSSI2016brochure,GSSI2021cart}, the 350~MHz 350HS antenna provides a typical detection depth of approximately 6~m and a maximum penetration of up to 12~m, depending on soil moisture and dielectric properties. 
The effective interpretation depth was empirically determined from the signal-to-noise ratio and the visibility of reflection hyperbolas in the processed B-scan images. 
In the collected field data, clear hyperbolic reflections were consistently observed up to about 3~m below the surface, beyond which the signals became indistinct due to increased attenuation in moist and conductive soils,
indicating that the radar performance was suitable for the experimental conditions of this study.

Both simulated and field B-scan datasets were processed using the same preprocessing pipeline to ensure consistent data quality. 
The procedures included background removal, automatic gain control (AGC), and band-pass filtering. 
Background removal was implemented by subtracting the median of 11 consecutive rows 
to eliminate static noise components. 
AGC was applied with a moving window of 151 rows and gain factors 
$G_{\mathrm{max}} = [1.0, 10, 10^2, 10^3, 10^4, 10^5]$ 
to enhance deeper reflections. 
A second-order band-pass filter (100–1500~MHz) was used to suppress low- and high-frequency noise. 
Table~\ref{tab:signal_proc} summarizes the input parameters used in the preprocessing steps. 
For simulated data, 1024 uniformly spaced time samples were selected 
to match the resolution of the field data.

\begin{table}[htbp]
\centering
\caption{Input parameters applied in signal processing.}
\label{tab:signal_proc}
\resizebox{\columnwidth}{!}{
    \begin{tabular}{lccc}
    \toprule
    \textbf{Process} & \textbf{Parameter} & \textbf{Symbol} & \textbf{Value} \\
    \midrule
    Background removal & Window size (row) & $w_\mathrm{bg}$ & 11 \\
    Auto Gain Control  & Gain factor & $g_{\max}$ & [1.0, 10, 10$^2$,\\ 
                       & & & 10$^3$, 10$^4$, 10$^5$] \\
                       & Window size (row) & $w_\mathrm{AGC}$ & 151 \\
    Band-pass filter   & High/Low cutoff (MHz) & $(f_H, f_L)$ & (1500, 100) \\
                       & Filter order & -- & 2 \\
    \bottomrule
    \end{tabular}
}
\end{table}

Table~\ref{tab:bscan_stats} summarizes the dataset characteristics per site, including image resolution, number of annotated objects, and average bounding box size. A total of 86 B-scan images were collected from Site~A, 70 from Site~B, and 11 from Site~C.

Although the number of real-world samples is limited (167 images in total, with 11 reserved for independent testing), 
statistical reliability was ensured using a five-fold cross-validation scheme across Sites~A and~B. 
In addition, multiple data augmentation techniques were applied in the YOLOv5 model to enhance sample diversity and prevent overfitting.

Although the simulated and real datasets have different original resolutions (1080×1080 and 680×680, respectively) and sample sizes (300 synthetic vs. 167 field images), 
all B-scan images were resized to 416×416 and underwent identical preprocessing and augmentation steps before training. 
This ensures consistent input dimensions and comparable learning conditions across all configurations.
Moreover, the simulated data were used solely for pretraining to learn generalizable reflection patterns, 
while the real datasets were used for fine-tuning and evaluation to assess domain transferability.

It should be noted that the simulated data were not generated to replicate any specific field site.
Instead, the simulation stage was designed to provide a controlled environment for learning fundamental GPR signal features—such as hyperbolic responses and subsurface reflections—under idealized conditions.
The real-world datasets used for training and testing were independently collected from different geographical locations (Sites~A, B, and~C), each exhibiting distinct soil compositions and noise characteristics.
Therefore, the proposed Sim2Real framework aims to transfer generalizable feature representations learned from simulation to heterogeneous real-world domains, rather than modeling a single site-specific scenario.

\begin{table}[thbp]
\centering
\caption{Descriptive statistics of B-scan datasets.}
\label{tab:bscan_stats}
\begin{tabularx}{\columnwidth}{l c c c}
\toprule
\textbf{Site} & \textbf{Res. (px)} & \textbf{\#Img/Box} & \textbf{Avg. Box (px)} \\
\midrule
Simulated & $1080\times1080$ & 300/300 & $280\!\pm\!33~\times~167\!\pm\!43$ \\
Site A    & $680\times680$   & 86/94   & $390\!\pm\!87~\times~107\!\pm\!23$ \\
Site B    & $680\times680$   & 70/70   & $234\!\pm\!53~\times~79\!\pm\!19$ \\
Site C    & $680\times680$   & 11/13   & $323\!\pm\!134~\times~129\!\pm\!38$ \\
\bottomrule
\end{tabularx}
\end{table}

\begin{table}[thbp]
\centering
\caption{Model configurations and dataset composition. Training and validation subsets are randomly divided from the combined data of Sites~A and~B,
while testing is performed exclusively on independent Site~C.}
\label{tab:models_config}
    \begin{tabularx}{\columnwidth}{l l l l}
    \toprule
    \textbf{Model} & \textbf{Train (Sim, A+B)} & \textbf{Validation (A+B)} & \textbf{Test (Site~C)} \\
    \midrule
    Base & $F_{\mathrm{tr}}$ & $F_{\mathrm{val}}$ & $F_{\mathrm{test}}$ \\
    Sim & $S_{\mathrm{total}}$ & $F_{\mathrm{val}}$ & $F_{\mathrm{test}}$ \\
    S2R & \makecell[l]{Pre-trained: $S_{\mathrm{total}}$ \\ Fine-tuned: $F_{\mathrm{tr}}$} & $F_{\mathrm{val}}$ & $F_{\mathrm{test}}$ \\
    TE-Base & $\widetilde{F}_{\mathrm{tr}}$ & $\widetilde{F}_{\mathrm{val}}$ & $\widetilde{F}_{\mathrm{test}}$ \\
    TE-Sim & $\widetilde{S}_\mathrm{total}$ & $\widetilde{F}_{\mathrm{val}}$ & $\widetilde{F}_{\mathrm{test}}$ \\
    \makecell[l]{TE-S2R \\ \textbf{(Proposed)}} & \makecell[l]{Pre-trained: $\widetilde{S}_\mathrm{total}$ \\ Fine-tuned: $\widetilde{F}_{\mathrm{tr}}$} & $\widetilde{F}_{\mathrm{val}}$ & $\widetilde{F}_{\mathrm{test}}$ \\
    \bottomrule
    \end{tabularx}
\vspace{3mm}
\footnotesize
\raggedright
\noindent
\textit{Note:} The simulated dataset contains 300 samples.
The field dataset includes 156 samples (125 for training, 31 for validation per fold),
and the test set consists of 11 independent samples from Site~C.
\end{table}

\subsection{Model Configurations}
To evaluate the effectiveness of the proposed shape-aware topological representation and the Sim2Real training strategy in object detection, we constructed six distinct model configurations, each designed to isolate the impact of sim2real, topological shape-aware representation, or their combination.

The simulated dataset consists of 300 synthetically generated samples and is used exclusively as a unified training set without partitioning. 
In contrast, the field dataset comprises 156 real-world samples collected from Site~A and Site~B, and is employed in a five-fold cross-validation scheme to ensure statistical robustness and consistency. Specifically, the field dataset is divided into five disjoint subsets $\{F^{(i)}\}_{i=1}^{5}$. 
For each fold, one subset $F^{(*)}$ is selected as the validation set, and the remaining four subsets are used for training:
\begin{equation}
F_{\mathrm{val}} = F^{(*)}, \quad F_{\mathrm{tr}} = \bigcup_{j \ne *} F^{(j)},
\end{equation}
where $* \in \{1, 2, 3, 4, 5\}$ denotes the fold index for the current validation set. 
These notations and data splits are consistently applied across all configurations to ensure fair comparison. 
In addition, 11 real-world samples from Site~C are employed as test dataset. 

We define six model configurations: \emph{Base}, \emph{Sim}, \emph{S2R}, \emph{TE-Base}, \emph{TE-Sim}, and \emph{TE-S2R}. 
Model~Base is trained solely on real field data $F_{\mathrm{tr}}$, while Model~Sim is trained exclusively on simulated data. 
Model~S2R adopts a domain adaptation approach by pre-training on simulated data and subsequently fine-tuning on field data, aiming to mitigate the domain gap between synthetic and real-world distributions.
Model~TE-Base, Model~TE-Sim, and Model~TE-S2R are the shape-aware counterparts of Model~Base, Model~Sim, and Model~S2R, respectively.
In the shape-aware configurations, all datasets—including training, validation, and test—are augmented with topological features extracted via persistent homology. 
These enhanced datasets are denoted with a tilde (e.g., $\widetilde{F}_{\mathrm{tr}}$, $\widetilde{\mathrm{Sim}}$) to indicate the inclusion of shape-aware structural descriptors.

Among all, Model~TE-S2R integrates both Sim2Real transfer learning and topological augmentation, and is regarded as the proposed model in this study due to its comprehensive design and superior performance across evaluation metrics.

Table~\ref{tab:models_config} summarizes the training, validation, and test dataset compositions used in each configuration, highlighting the combinations of data sources and the presence or absence of topological augmentation.
For clarity, we denote the real-world field dataset by $F$, with subsets $F_{\mathrm{tr}}$, $F_{\mathrm{val}}$, and $F_{\mathrm{test}}$ representing the training, validation, and test splits, respectively. 
Similarly, $S_{\mathrm{total}}$ refers to the full set of simulated samples used for training-only purposes. 
Tilde symbols (e.g., $\widetilde{F}_{\mathrm{tr}}$, $\widetilde{S}_{\mathrm{total}}$) indicate that the corresponding data have been augmented with topological features extracted via persistent homology.

For all model configurations, the dataset is divided into three disjoint subsets—training, validation, and testing.
\begin{itemize}{
    \item Model~Base and Model~TE-Base, the training and validation subsets are drawn from the combined field data of Sites~A and~B, comprising 125 samples for training and 31 samples for validation in each fold, following a five-fold cross-validation scheme to ensure statistical robustness.
        The folds were stratified such that samples from both sites were evenly distributed across all folds, thereby avoiding any site-specific bias in training or validation.
    \item Model~S2R and Model~TE-S2R are first pre-trained using 300 simulated B-scan images and subsequently fine-tuned with the same training and validation protocol as Model~Base and Model~TE-Base.
    \item Model~Sim and Model~TE-Sim are trained solely on the 300 simulated B-scans but evaluated using the same five-fold validation process for consistency.}
\end{itemize}
All configurations are finally tested on the independent Site~C dataset (11 samples), which is completely excluded from training and fine-tuning, providing a strict evaluation of generalization performance under domain-shift conditions.

\subsection{Training Setup}
We employed the YOLOv5s architecture, a lightweight and computationally efficient variant of the YOLOv5 family with approximately 7 million parameters.
Its fast inference speed and reduced memory footprint make it suitable for GPR-based object detection where real-time performance and hardware efficiency are desired.
All models were trained using consistent hyperparameters based on the YOLOv5s architecture.
All input images were resized to $416 \times 416$ to comply with the model's fixed input resolution.
The training process differs by configuration: 

For the Sim2Real transfer-learning procedure, the YOLOv5s model was first pre-trained on 300 simulated B-scan images using standard detection losses 
(objectness, classification, and bounding-box regression).
The learned weights from this pre-training stage were then transferred to the fine-tuning stage using real field data (Sites~A and~B), while the independent Site~C dataset was reserved exclusively for testing. 
This process enables the model to retain fundamental reflection patterns learned from simulation and adapt them to domain-specific variations in real environments.

For training or pre-training, we employed the following hyperparameters:
\begin{itemize}
    \item \textbf{Model config}: yolov5s.yaml
    \item \textbf{Pre-trained weights}: yolo5s.pt
    \item \textbf{Input resolution}: $416 \times 416$ pixels
    \item \textbf{Batch size}: 16
    \item \textbf{Training epochs}: 100
    \item \textbf{Optimizer}: SGD with momentum = 0.937, weight decay = 0.0005
    \item \textbf{Learning rate}: Linearly decayed from 0.01 to 0.0001
    \item \textbf{Warmup}: 3-epoch linear warmup increasing learning rate (from 0 to 0.01) and momentum (from 0.8 to 0.937); bias LR starts at 0.1
    \item \textbf{Data augmentation}: Mosaic (probability = 1.0), random horizontal flipping (probability = 0.5), HSV augmentation (hue = 0.015, saturation = 0.7, value = 0.4), translation (0.1), scaling (0.5)
    \item \textbf{Loss function}: Binary cross entropy for classification and objectness, IoU loss for box regression (box = 0.05, cls = 0.6, iou\_t = 0.2)
    \item \textbf{Anchor optimization}: Auto-anchor enabled with anchor\_t = 4.0 for optimal anchor matching
\end{itemize}

For fine-tuning, we initialized the model with pre-trained weights and applied the following settings:
\begin{itemize}
    \item \textbf{Model config}: yolov5s.yaml
    \item \textbf{Pre-trained weights}: Pre-trained weights from simulated GPR data (\emph{best.pt})
    \item \textbf{Optimizer}: SGD with momentum = 0.937, weight decay = 0.0005
    \item \textbf{Learning rate}: Linearly decayed from 0.01 to 0.0001
    \item \textbf{Batch size}: 16
    \item \textbf{Input resolution}: $416 \times 416$ pixels
    \item \textbf{Training epochs}: 100
    \item \textbf{Frozen layers}: None (all layers were trainable) 
    \item \textbf{Data augmentation}: Mosaic (probability = 1.0), random horizontal flipping (probability = 0.5), HSV augmentation (hue = 0.015, saturation = 0.7, value = 0.4), translation (0.1), scaling (0.5)
\end{itemize}

Although fine-tuning commonly involves smaller learning rates, shorter training durations and several frozen layers, we observed that retaining the original learning schedule yielded more stable convergence and better generalization in our case, possibly due to the domain gap between simulated and field data, as well as the strong regularization effects from extensive data augmentation.
All experiments were conducted using PyTorch 2.7.1 with CUDA 12.7 on a workstation equipped with a single NVIDIA GeForce RTX 4060 GPU (8~GB VRAM).
\begin{table*}[htbp]
\centering
\caption{YOLOv5-based performance comparison across six model configurations using 5-fold cross-validation (mean ± std). 
Results are reported separately for validation and test sets.}
\label{tab:yolo_v5_final_results}
\resizebox{\textwidth}{!}{
    \begin{tabular}{c|cccccccc}
    \toprule
    \textbf{Model} & \makecell{\textbf{mAP@0.5} \\ \textbf{(Val)}} & \makecell{\textbf{mAP@0.5} \\ \textbf{(Test)}} & \makecell{\textbf{mAP@0.5:0.95} \\ \textbf{(Val)}} & \makecell{\textbf{mAP@0.5:0.95} \\ \textbf{(Test)}} & \makecell{\textbf{Precision} \\\textbf{(Val)}} & \makecell{\textbf{Precision} \\ \textbf{(Test)}} & \makecell{\textbf{Recall} \\ \textbf{(Val)}} & \makecell{\textbf{Recall} \\ \textbf{(Test)}} \\
    \midrule 
    Base & 0.986±0.009 & 0.357±0.089 & 0.677±0.031 & 0.115±0.028 & 0.9649±0.037 & 0.787±0.220 & 0.9774±0.0201 & 0.293±0.035 \\
    Sim & 0.276±0.087 & 0.031±0.037 & 0.138±0.054 & 0.013±0.015 & 0.311±0.070 & 0.024±0.030 & 0.204±0.070 & 0.062±0.069 \\
    S2R & 0.992±0.003 & 0.591±0.097 & 0.693±0.007 & 0.274±0.025 & 0.970±0.016 & 0.894±0.080 & 0.993±0.014 & 0.374±0.126 \\
    TE-Base & 0.985±0.010 & 0.494±0.090 & 0.559±0.049 & 0.190±0.016 & 0.984±0.020 & 0.917±0.156 & 0.963±0.024 & 0.410±0.110 \\
    TE-Sim & 0.386±0.119 & 0.110±0.065 & 0.130±0.021 & 0.051±0.029 & 0.411±0.281 & 0.387±0.410 & 0.378±0.147 & 0.108±0.037 \\
    \textbf{TE-S2R} & 0.990±0.003 & \textbf{0.643±0.069} & 0.704±0.012 & \textbf{0.306±0.046} & 0.967±0.022 & \textbf{0.918±0.054} & 0.988±0.025 & \textbf{0.456±0.057} \\
    \end{tabular}
}
\end{table*}

\begin{table}[htbp]
\centering
\caption{YOLOv11-based performance comparison across six model configurations using 5-fold cross-validation (mean ± std). 
Only test-set results are reported to verify consistency across architectures.}
\label{tab:yolov11_final_results}
\begin{tabular}{c|cccc}
\toprule
\textbf{Model} & \makecell{\textbf{mAP@0.5} \\ \textbf{(Test)}} & \makecell{\textbf{mAP@0.5:0.95} \\ \textbf{(Test)}} & \makecell{\textbf{Precision} \\ \textbf{(Test)}} & \makecell{\textbf{Recall} \\ \textbf{(Test)}} \\
\midrule
Base & 0.773±0.108 & 0.326±0.040 & 0.845±0.126 & 0.65±0.067\\
Sim & 0.040±0.089 & 0.016±0.036 & 0.050±0.112 & 0.017±0.037 \\
S2R & 0.795±0.094 & 0.353±0.087 & 0.873±0.086 & 0.703±0.088 \\
TE-Base & 0.771±0.046 & 0.360±0.040 & 0.894±0.061 & 0.714±0.050\\
TE-Sim  & 0.150±0.088 & 0.024±0.033 & 0.080±0.108 & 0.120±0.025\\
\textbf{TE-S2R} & \textbf{0.806±0.016} & \textbf{0.378±0.024}  & \textbf{0.934±0.052} & \textbf{0.748±0.004}\\
\end{tabular}
\end{table}

\subsection{Results and Comparison}

To evaluate detection performance, we compute standard object detection metrics: \emph{Precision}, \emph{Recall}, \emph{mAP@0.5}, and \emph{mAP@0.5:0.95} on validation and test set.
\emph{Precision} and~\emph{Recall} are used to evaluate the binary detection task of identifying buried pipes.
\emph{Precision} represents the proportion of predicted boxes that correctly match actual buried targets, while~\emph{Recall} quantifies the proportion of true buried objects that are successfully detected. 

Ground truth bounding boxes are generated differently depending on the data source. For simulated data, exact object positions are predefined in the gprMax configuration. 
For field data, ground truth annotations are based on measured burial depths and object dimensions recorded on the drawing. 
These annotations serve as ground truth for evaluating detection performance.

Tables~\ref{tab:yolo_v5_final_results} provides the final 5-fold cross-validation results on the validation and test sets. 
These results allow for an in-depth evaluation of each configuration's generalization capability under domain shift conditions.

Model~Sim and Model~TE-Sim, which are trained solely on simulated data, exhibit significantly degraded performance on the test set compared to the validation set. 
This decline highlights a substantial domain gap between synthetic and real-world data, and underscores the limited generalization ability of models trained exclusively on simulation.

\begin{figure}[thbp]
    \centering
    \includegraphics[width=0.6\columnwidth]{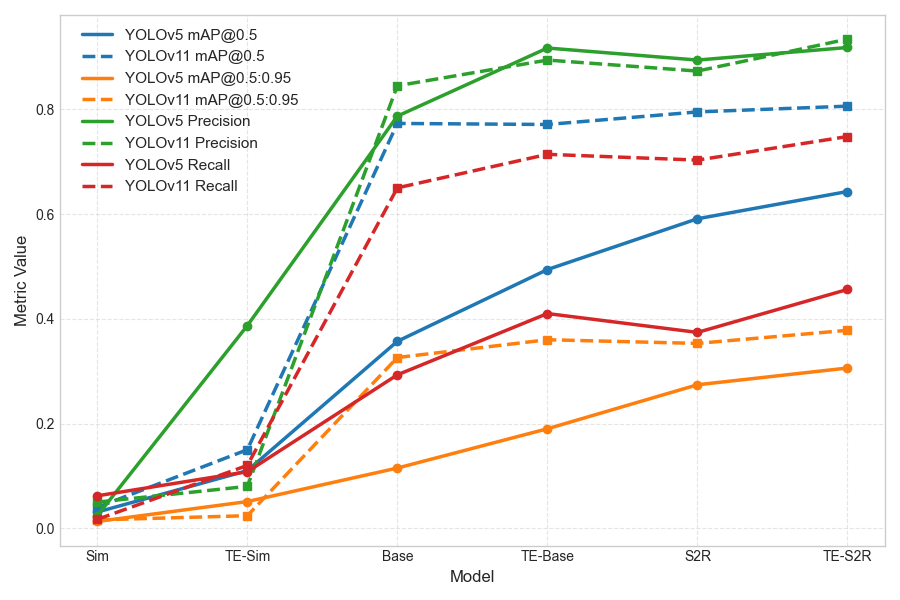}
    \caption{
    Line chart comparing the test-set performance of YOLOv5- and YOLOv11-based models across six configurations.
    }
    \label{fig:linechart_model_comparison}
\end{figure}

To complement these numerical results, Fig. ~\ref{fig:linechart_model_comparison} visualizes the overall performance trends across all six configurations. 
The line chart clearly illustrates that model performance improves progressively from \textit{Sim} to \textit{TE-S2R}, 
demonstrating the consistent advantage of incorporating topological enhancement (TDA) and Sim2Real transfer learning. 
In particular, the proposed TE-S2R model achieves the highest Precision, Recall, and mAP values, confirming its superior accuracy and robustness under domain-shift conditions.

To further validate the robustness contribution of the proposed topological features, additional experiments were conducted using the YOLOv11 detector. 
To complement the numerical results, Fig. ~\ref{fig:linechart_model_comparison} compares the test-set performance trends of YOLOv5 and YOLOv11 across six configurations. 
This line chart highlights the consistency of model behavior between the two architectures: 
both exhibit progressive improvement from \textit{Base} to \textit{TE-S2R}, demonstrating the persistent benefit of incorporating topological enhancement (TDA) and Sim2Real transfer learning. 
In particular, the proposed TE-S2R configuration consistently achieves the highest Precision, Recall, and mAP values in both detectors, confirming its superior accuracy and robustness under domain-shift conditions.

The decrease in mAP and Recall on the test set mainly arises from the domain gap between the validation (Sites~A and~B) and test (Site~C) datasets, 
which clearly demonstrates the independence of the test set. 
Nevertheless, Precision remains relatively stable, indicating that the model preserves discriminative reliability despite increased environmental variability. 
Across both YOLOv5 and YOLOv11 architectures, the TDA-enhanced configurations (TE-Base and TE-S2R) consistently exhibited higher stability and accuracy under noisy and heterogeneous soil conditions, particularly on the independent test dataset, further confirming that persistent homology improves global feature consistency and noise resistance.

\begin{figure}[thbp]
    \centering
    \subfloat[Ablation study on validation dataset]{
        \includegraphics[width=\linewidth]{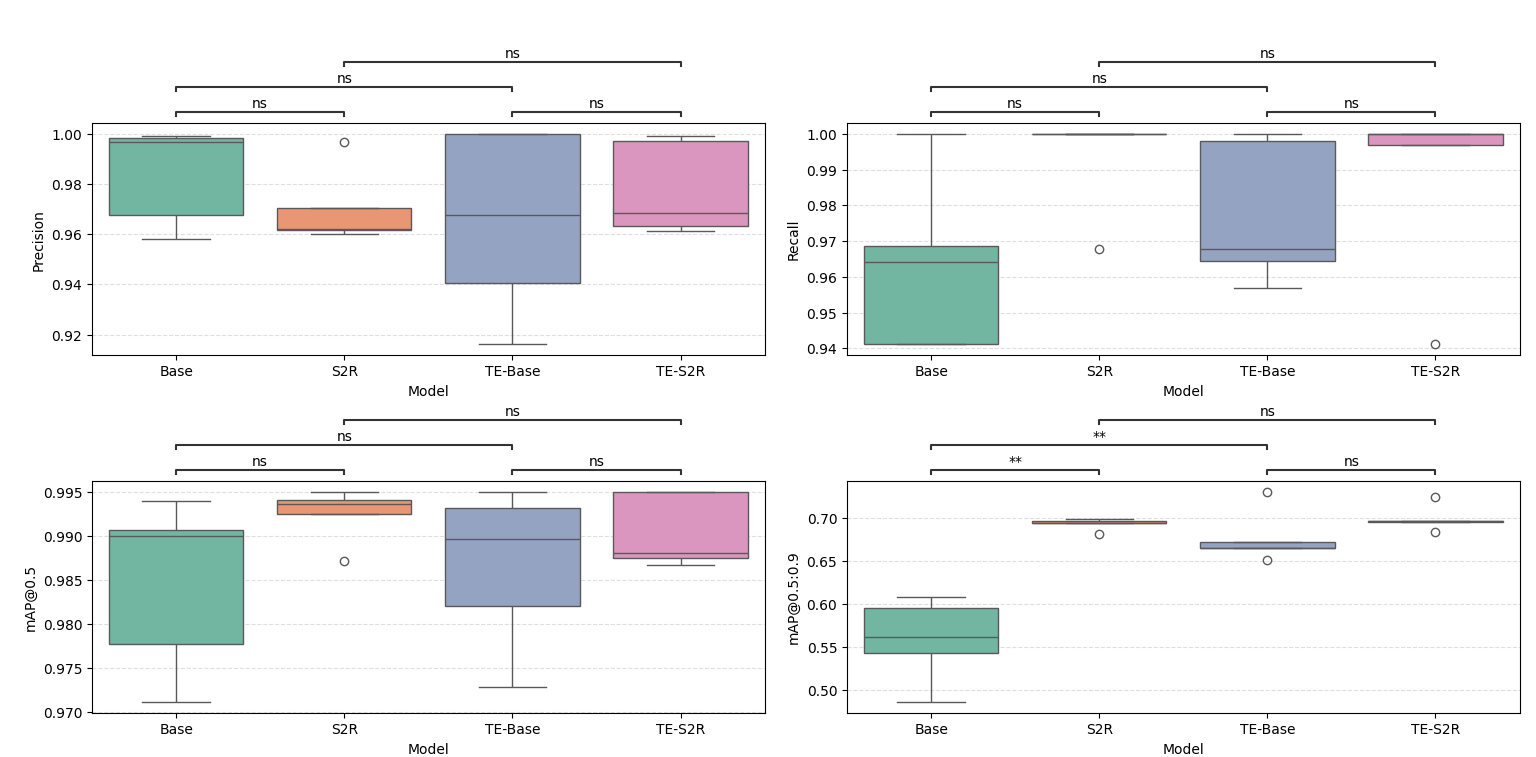}
    }
    \hfill
    \subfloat[Ablation study on test dataset]{
        \includegraphics[width=\linewidth]{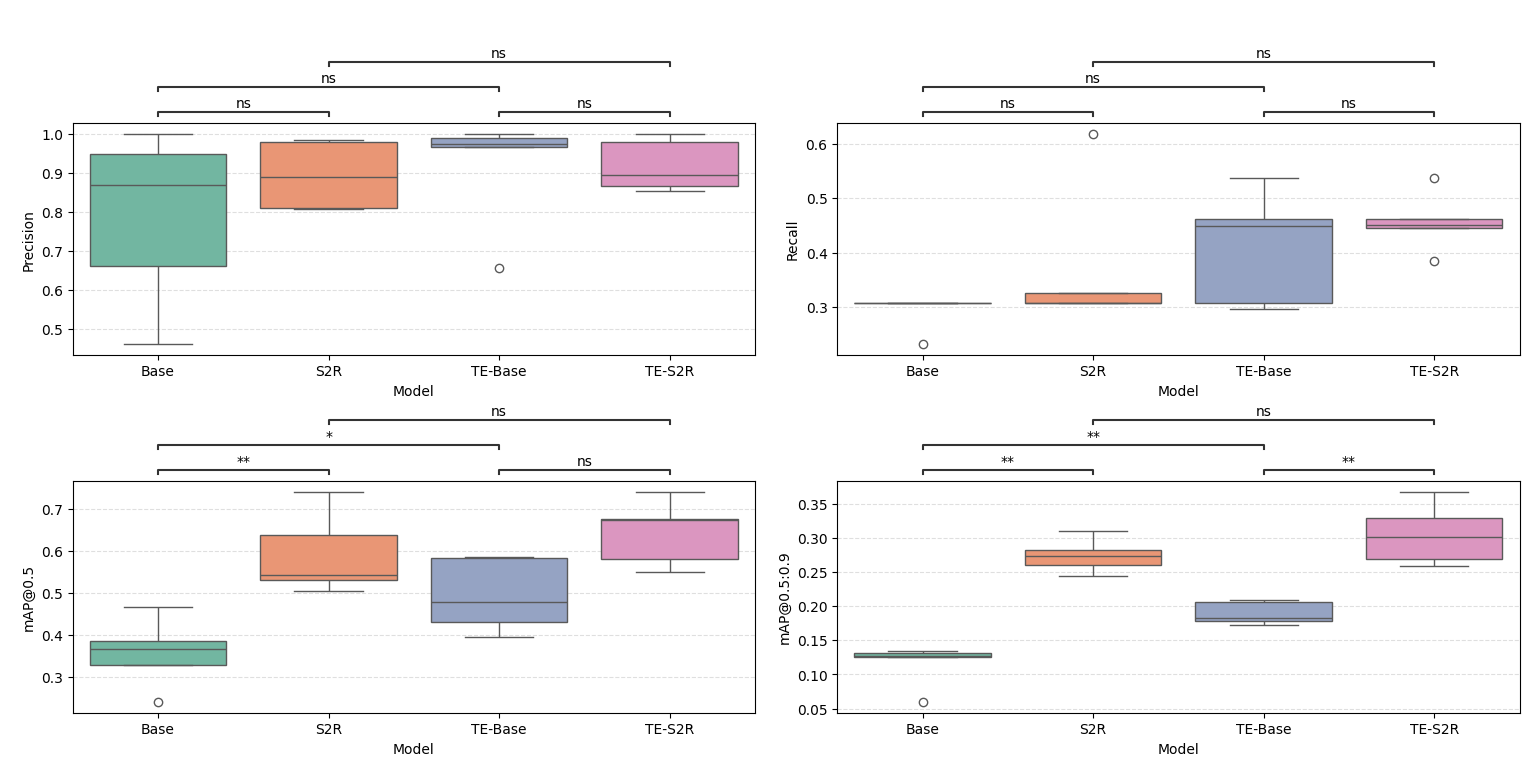}
    } 
    \caption{
    Ablation study evaluating the individual contributions of Sim2Real transfer learning and shape-aware topological augmentation (TDA). 
    Models (a)--(f) correspond to the configurations in Table~\ref{tab:models_config}. 
    Mann--Whitney U test was applied to assess statistical significance. 
    Annotations indicate: ‘**’ for $p < 0.01$, ‘*’ for $p < 0.05$, and ‘ns’ for $p \geq 0.05$.
    }
    \label{fig:ablation_study}   
\end{figure}

To better understand the individual contributions of the Sim2Real strategy and the proposed shape-aware topological representation, we conducted an ablation study focusing on Model~\emph{Base}, Model~S2R, Model~TE-Base, and Model~TE-S2R. The results are illustrated in Fig.  ~\ref{fig:ablation_study}~(a) and Fig. ~\ref{fig:ablation_study}~(b), corresponding to the validation and test sets, respectively.

Across both validation and test results, we observe two consistent trends. 
First, comparisons between Model~Base and Model~S2R, as well as Model~TE-Base and Model~TE-S2R, highlight the effectiveness of Sim2Real transfer learning. 
Second, performance improvements from Model~Base to Model~TE-Base, and from Model~S2R to Model~TE-S2R, demonstrate the benefit of incorporating shape-aware features via TDA.
We observe that TDA-enhanced configurations generally outperform their non-TDA counterparts, particularly under domain shift conditions.
This robustness arises because topological features capture global structural patterns that remain stable, even when local pixel-level intensities vary due to noise, material properties, or sensor inconsistencies.

Although the Base model trained purely on real data shows competitive results on the training sites (A and B),
its performance does not generalize well to unseen domains due to the limited dataset size and site-specific conditions. 
The proposed Sim2Real and TDA-enhanced configurations mitigate this issue by pretraining on diverse simulated scenarios and transferring learned representations to real environments.
This strategy significantly improves robustness and generalization, particularly on the independent Site C dataset.

To assess the statistical significance of these observations, we applied the Mann–Whitney U test to all pairwise comparisons between configurations. The resulting $p$-values are annotated in the figures using standard notation, where ‘**’ denotes $p < 0.01$ (statistically significant), and ‘ns’ indicates no significant difference ($p \geq 0.05$).

Among all configurations, Model~TE-S2R which integrates both Sim2Real and shape-aware topological representation—consistently outperforms the others across all evaluation metrics. 
This confirms the effectiveness of the proposed approach in improving both accuracy and robustness, particularly under domain shift conditions. 
Notably, the TE-S2R model achieves higher Recall, mAP@0.5, and mAP@0.5:0.95 scores even when fine-tuned with the same limited amount of annotated field data, and further demonstrates improved generalization performance on the completely independent test set (Site~C).

\begin{figure*}[thbp]
    \centering
    \subfloat[Label]{
        \includegraphics[width=0.6\textwidth]{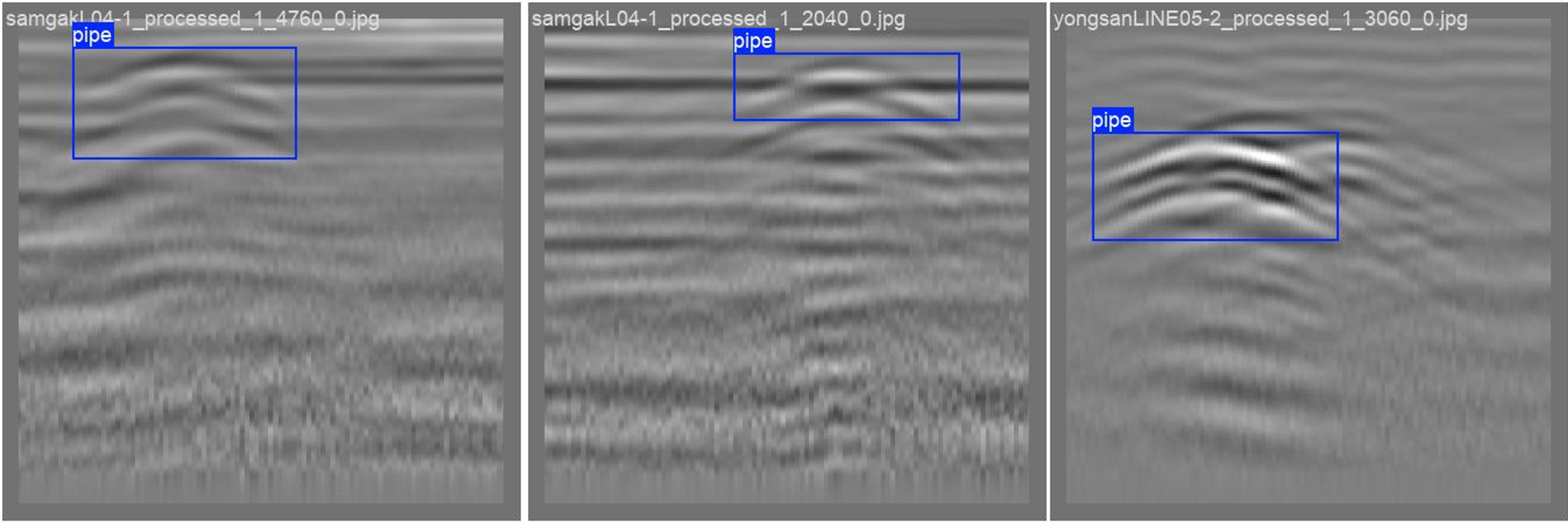}
    }
    
    \subfloat[Base]{
        \includegraphics[width=0.49\textwidth]{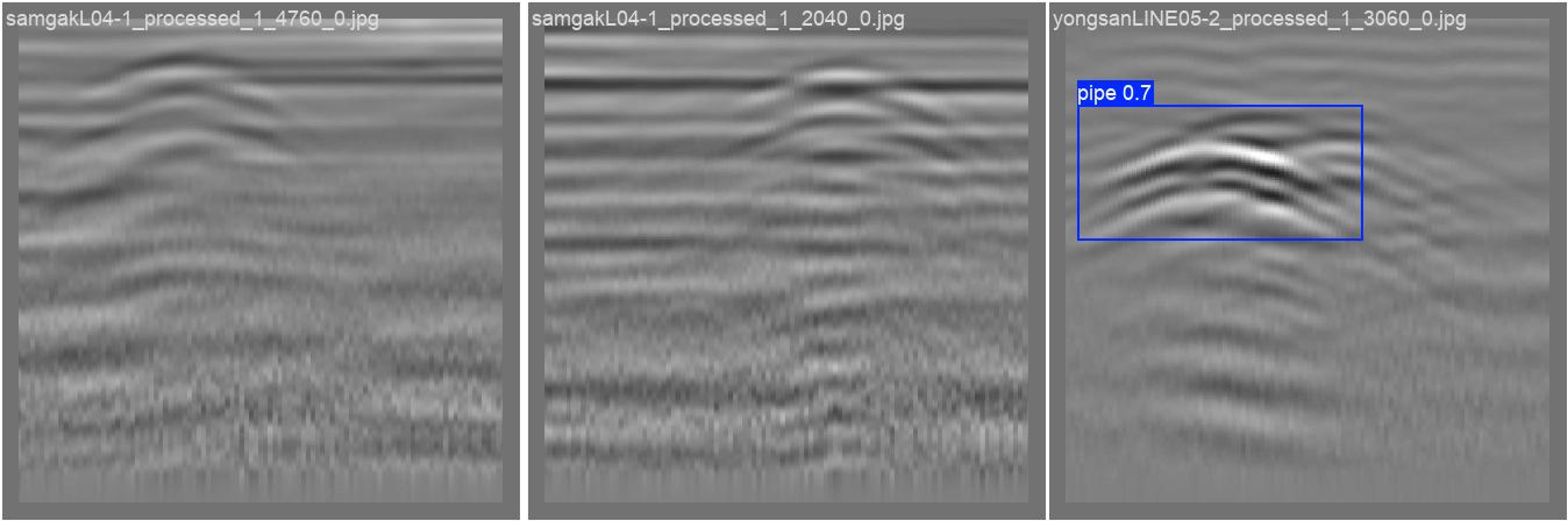}
    }
    \subfloat[TE-Base]{
        \includegraphics[width=0.49\textwidth]{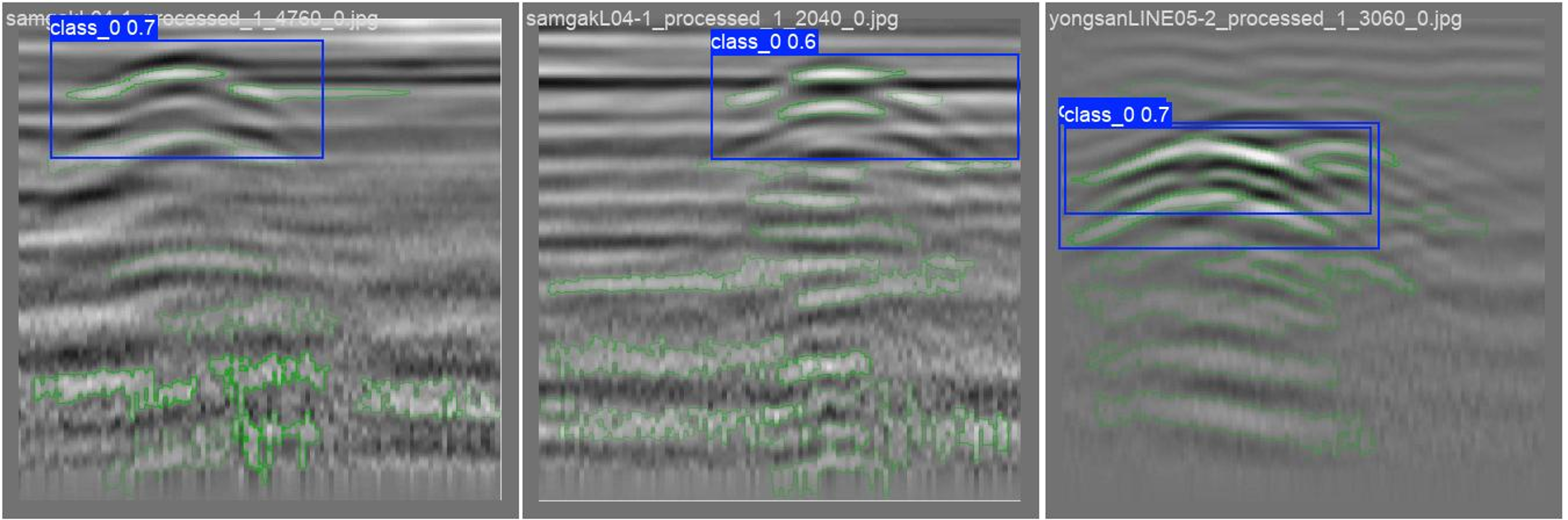}
    }
    
    \subfloat[S2R]{
        \includegraphics[width=0.49\textwidth]{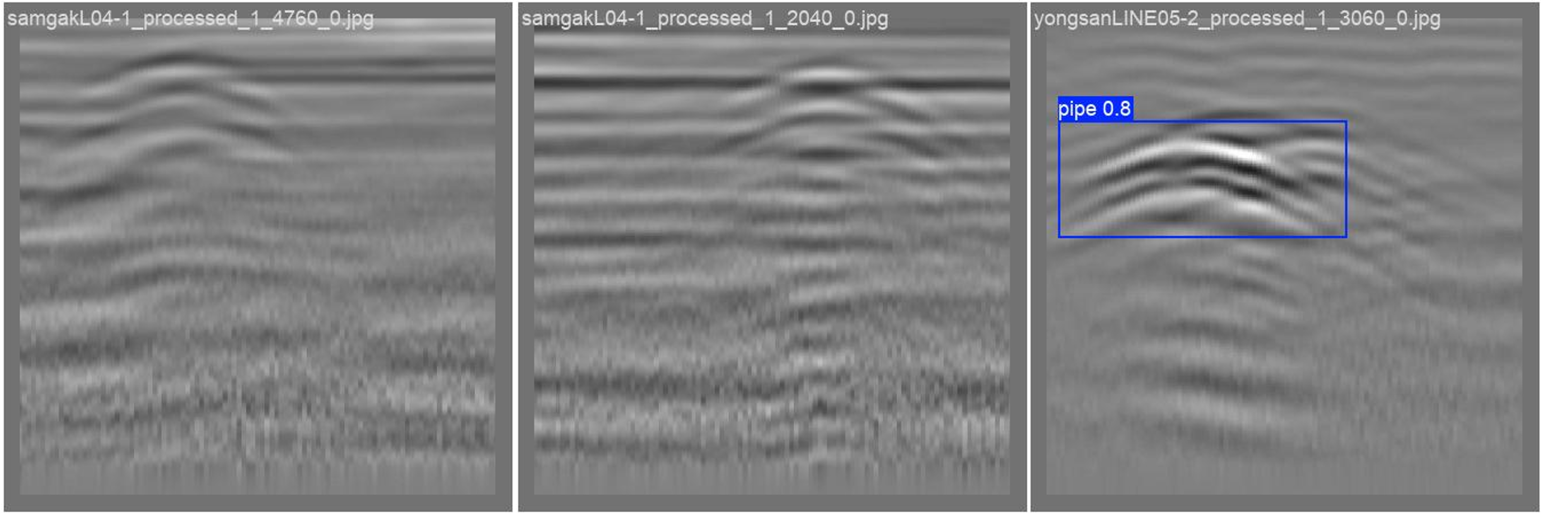}
    }
    \subfloat[TE-S2R]{
        \includegraphics[width=0.49\textwidth]{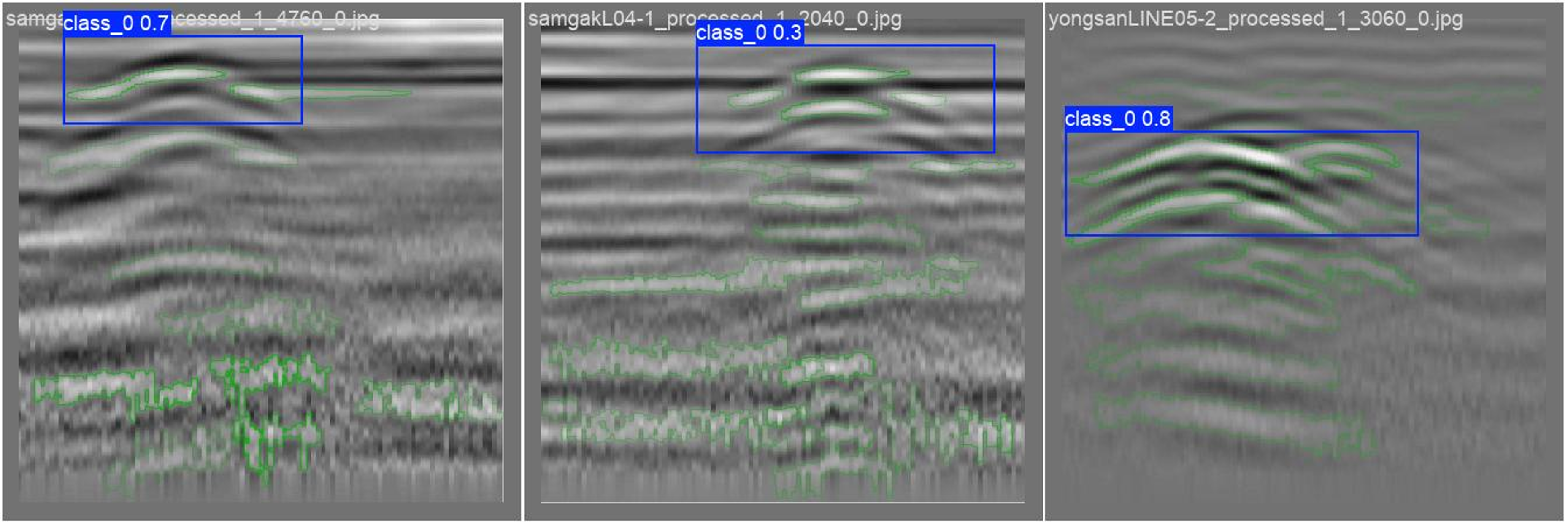}
    }
    \caption{
    Qualitative comparison of detection results across representative configurations on a real B-scan image from the test site (Site~C). 
    Each greed bounding box indicates a detected buried pipe.
    }
    \label{fig:qualitative_comparison}
\end{figure*}

To visually demonstrate the effect of topological enhancement, 
Fig.~\ref{fig:qualitative_comparison} presents representative detection results 
on a real B-scan image from the independent test site (Site~C).
Compared with the Base and S2R configurations, 
the TDA-enhanced models (TE-Base and TE-S2R) produce clearer hyperbolic reflections and fewer false detections, 
visually confirming the quantitative improvements discussed above.

\section*{Conclusion}
This paper proposed a novel framework, \emph{TE-S2R} (TDA-Enhanced Sim2Real), that integrates shape-aware topological representations with deep learning-based object detection for GPR data. By incorporating lifetime-weighted persistent homology into the input representation, the proposed method enhances sensitivity to buried object signatures that are often difficult to detect in conventional B-scan images. To the best of our knowledge, this is the first study to directly embed homological information as shape descriptors within a deep learning pipeline for GPR analysis, representing a meaningful departure from purely visual approaches.

A key contribution of this work lies in the synergistic integration of Topological Data Analysis (TDA) and Sim2Real (S2R) strategies. While S2R methods often suffer from domain gaps caused by structural inconsistencies and noise between synthetic and real-world data, TDA mitigates this issue by extracting global structural features that remain stable across domains. These topological features are inherently robust to local perturbations and help align the shape characteristics between simulated and field data. In TE-S2R, TDA not only strengthens the model's resistance to noise but also enables a shared topological understanding across domains—transforming the S2R process from a simple domain adaptation step into a structure-aware transfer mechanism.

Experimental results demonstrate that TE-S2R consistently outperforms baseline models in terms of mean Average Precision (mAP), recall, and precision. The combination of topological priors and S2R enhances both generalization and robustness, enabling reliable detection even under real-world GPR conditions. These findings indicate that the integration of TDA and S2R is not merely additive but mutually reinforcing, yielding a performance gain that exceeds what either method can achieve alone.

Overall, the TE-S2R framework offers a scalable and noise-resistant solution for subsurface sensing tasks and can be readily extended beyond GPR to other imaging modalities where structural information is crucial but often obscured by noise or signal distortion.

Future work will focus on extending the framework to multi-class scenarios, exploring self-supervised topological learning, and validating performance in broader geophysical environments.

\end{document}